\documentclass{article}

\usepackage{setspace}
\usepackage{color}
\usepackage{listings}
\usepackage{graphicx}
\usepackage[margin=1in]{geometry}
\usepackage{breakcites}
\usepackage[bottom]{footmisc}
\usepackage{float}
\usepackage[numbers]{natbib}
\usepackage{booktabs}
\usepackage{multirow}
\usepackage{siunitx}
\usepackage{adjustbox}
\makeatletter
\renewcommand\@makefntext[1]{\leftskip=2em\hskip-1em\@makefnmark#1}
\makeatother
\usepackage{array}
\newcolumntype{L}[1]{>{\raggedright\let\newline\\\arraybackslash\hspace{0pt}}m{#1}}
\newcolumntype{C}[1]{>{\centering\let\newline\\\arraybackslash\hspace{0pt}}m{#1}}
\newcolumntype{R}[1]{>{\raggedleft\let\newline\\\arraybackslash\hspace{0pt}}m{#1}}

\usepackage{tabularx}
\usepackage{hyperref}
\usepackage{graphicx}% Include figure files
\usepackage{dcolumn}% Align table columns on decimal point
\usepackage{bm}% bold math
\usepackage{amsmath}
\usepackage{amsfonts}
\usepackage{amssymb}

\definecolor{orange}{RGB}{255,127,0}
\definecolor{purple}{RGB}{128,0,128}

\pdfoutput=1

\usepackage{authblk}
\title{Genetic and neuroanatomical support for functional brain network dynamics in epilepsy}

\author[1]{Pranav G. Reddy}
\author[1,2,3]{Richard F. Betzel}
\author[4]{Ankit N. Khambhati}
\author[1]{Preya Shah}
\author[1]{Lohith Kini}
\author[1,5]{Brian Litt}
\author[6]{Timothy H. Lucas}
\author[1,5]{Kathryn A. Davis}
\author[1,5,7,8,*]{Danielle S. Bassett}
\affil[1]{Department of Bioengineering, School of Engineering and Applied Sciences, University of Pennsylvania, Philadelphia, PA 19104 USA}
\affil[2]{Department of Psychological \& Brain Sciences, Indiana University, Bloomington, IN 47405 USA}
\affil[3]{Cognitive Science Program, Indiana University, Bloomington IN 47405 USA}
\affil[4]{Department of Neurological Surgery, University of California, San Francisco, CA 94122 USA}
\affil[5]{Department of Neurology, Perelman School of Medicine, University of Pennsylvania, Philadelphia, PA 19104 USA}
\affil[6]{Department of Neurosurgery, Perelman School of Medicine, University of Pennsylvania, Philadelphia, PA 19104 USA}
\affil[7]{Department of Electrical \& Systems Engineering, School of Engineering and Applied Sciences, University of Pennsylvania, Philadelphia, PA 19104 USA}
\affil[8]{Department of Physics \& Astronomy, School of Arts \& Sciences, University of Pennsylvania, Philadelphia, PA 19104 USA}
\affil[*]{To whom correspondence should be addressed: dsb@seas.upenn.edu}

\date{}                     %% if you don't need date to appear
\begin{document}

\maketitle

\section*{Abstract}
Focal epilepsy is a devastating neurological disorder that affects an overwhelming number of patients worldwide, many of whom prove resistant to medication. The efficacy of current innovative technologies for the treatment of these patients has been stalled by the lack of accurate and effective methods to fuse multimodal neuroimaging data to map anatomical targets driving seizure dynamics. Here we propose a parsimonious model that explains how large-scale anatomical networks and shared genetic constraints shape inter-regional communication in focal epilepsy. In extensive ECoG recordings acquired from a group of patients with medically refractory focal-onset epilepsy, we find that ictal and preictal functional brain network dynamics can be accurately predicted from features of brain anatomy and geometry, patterns of white matter connectivity, and constraints complicit in patterns of gene coexpression, all of which are conserved across healthy adult populations. Moreover, we uncover evidence that markers of non-conserved architecture, potentially driven by idiosyncratic pathology of single subjects, are most prevalent in high frequency ictal dynamics and low frequency preictal dynamics. Finally, we find that ictal dynamics are better predicted by white matter features and more poorly predicted by geometry and genetic constraints than preictal dynamics, suggesting that the functional brain network dynamics manifest in seizures rely on -- and may directly propagate along -- underlying white matter structure that is largely conserved across humans. Broadly, our work offers insights into the generic architectural principles of the human brain that impact seizure dynamics, and could be extended to further our understanding, models, and predictions of subject-level pathology and response to intervention.

\newpage
\section*{Introduction}

\noindent For over 60 million patients, epilepsy restricts the quality of daily life through spontaneous and recurring seizures. While seizures may be controlled in two-thirds of epilepsy patients, the remaining require more invasive treatment. Medication-resistant patients undergo continuous monitoring of intracranial electrophysiology for biomarkers generated by the epileptic network, a set of interacting brain regions that are thought to initiate and spread seizure activity in the brain \citep{siegel2001medically}. Novel technologies have expanded treatment options beyond surgical resection of abnormal tissue to laser ablation and neurostimulation \citep{stacey2008technology, fisher2010electrical, morrell2011responsive, tovar2013use, medvid2015current}, affording greater specificity in targeting discrete nodes of a patient's epileptic network. To optimize an intervention strategy for seizure control, practitioners are required to assimilate neuroimaging data across multiple modalities, and to map anatomical targets where intervention would most likely reduce seizure dynamics. However, a parsimonious mechanism explaining how large-scale anatomical networks shape inter-regional communication in focal epilepsy has remained elusive. Understanding how large-scale brain architecture facilitates the onset and rapid spread of seizures can yield better tools for mapping targets for therapy that broadly extend to the patient cohort.\\

\noindent A large body of research has shown that spatially distributed variations in anatomy can distinguish patients with epilepsy from healthy individuals \citep{gross2006extratemporal, riley2010altered, otte2012meta}. Quantitative features of white-matter connectivity can also explain first-order parameters of seizure dynamics, such as duration \citep{chiang2016white} and seizure severity \citep{labate2015white}. Astonishingly detailed seizure dynamics can be predicted using \textit{in silico} network modelling that fuses a patient's structural connectome -- a comprehensive network map of physical connections among neural elements -- with their intracranial electrophysiology recordings \citep{proix2017individual, jirsa2017virtual}. While variation in anatomy and pathophysiology can precipitate seizures through different mechanisms \citep{frauscher2017different, salami2015distinct}, seizures also exhibit substantial similarities in their dynamics within and across patients \citep{khambhati2015dynamic, burns2014network, jacobs2008interictal, wilke2011graph}. Indeed, these and related studies suggest that certain properties of the epileptic network may be common across patients. A critical question, then, is to what extent are seizure dynamics a product of anatomical organization that is common across individuals and more fundamental to the human brain?\\

\noindent Recent work has demonstrated a significant convergence of structural network features across large populations of healthy individuals \citep{bassett2011conserved,betzel2016generative}, spanning both young and old age \citep{zuo2017human,baum2017modular}, and extending even to diverse patient groups with neurological disease \citep{stam2014modern} and psychiatric disorders \citep{fornito2015connectomics,park2018neuroanatomical}. Notably, such conserved structural features appear to be reflected in patterns of gene coexpression in both the mouse \citep{betzel2018diversity} and human \citep{betzel2017inter,romero2018structural}. The modular pattern of gene expression is consistent across independent human datasets, evolutionarily conserved in non-human primates \citep{anderson2018gene} and other mammals \citep{fulcher2016transcriptional}, and supports synchronous activity in brain networks \citep{richiardi2015correlated,vertes2016gene}. Collectively, these findings suggest that gene expression is an important marker of conserved structural features of human brain networks, and can by extension also reflect functional dynamics in distributed neuronal ensembles. It is therefore of interest to ask whether and how the patterns of structural connectivity and gene coexpression that are conserved across healthy individuals might offer principles upon which the pathology of refractory epilepsy depends.\\

\noindent To answer this question, we constructed dynamic functional networks using electrocorticographic (ECoG) neural recordings from human epilepsy patients undergoing routine clinical evaluation for epilepsy surgery. Recent work has addressed several challenges in studying ECoG dynamics across multiple individuals by accounting for variable electrode placement and sparse cortical coverage, developing a method to study how resting state ECoG functional networks can be predicted from underlying white matter tractography, spatial features, and gene coexpression \citep{betzel2017inter}, supporting related work in non-human species \citep{mills2018correlated}. Here, we extend this approach with an expanded set of structural features to model and predict ictal ECoG functional networks to measure the relative impact of different communication policies over white matter networks on epileptic seizures. This method is a unique aggregation of data across multiple approaches to the study of epilepsy, including gene coexpression data from the Allen Institute \citep{hawrylycz2012anatomically} and structural connectivity measures from diffusion imaging combined within a principled machine learning framework. We hypothesize that models trained with structural parameters would best predict functional connectivity during seizures, thus suggesting the influence of specific communication policies and measures over white matter in explaining ictal functional connectivity.\\

\noindent To test this hypothesis, we recorded ECoG from 25 patients diagnosed with drug-resistant epilepsy undergoing routine pre-surgical evaluation. We separated these recordings into pre-seizure and seizure epochs and constructed ECoG functional networks for both of them separately. The seizure epoch spanned the period between the clinically-marked earliest electrographic change and seizure termination, while the pre-seizure epoch was identical in duration to the seizure and ended immediately prior to the earliest electrographic change. In each epoch, we divided the ECoG signal into 1 sec non-overlapping time-windows and estimated functional connectivity in $\alpha/\theta$ (5--15 Hz), $\beta$ (15--25 Hz), low-$\gamma$ (30--40 Hz), and high-$\gamma$ (95--105 Hz) frequency bands using multitaper coherence estimation. We trained multi-linear regression models with different combinations of structural, physical, and genetic parameters in different settings to determine the genetic and neuroanatomical mechanisms predictive of ictal connectivity.

\section*{Results}

\subsection*{Preictal and ictal functional connectivity are generically predicted by structure, geometry, and genetics}

\begin{figure}[H]
\centering
\includegraphics[width=150mm]{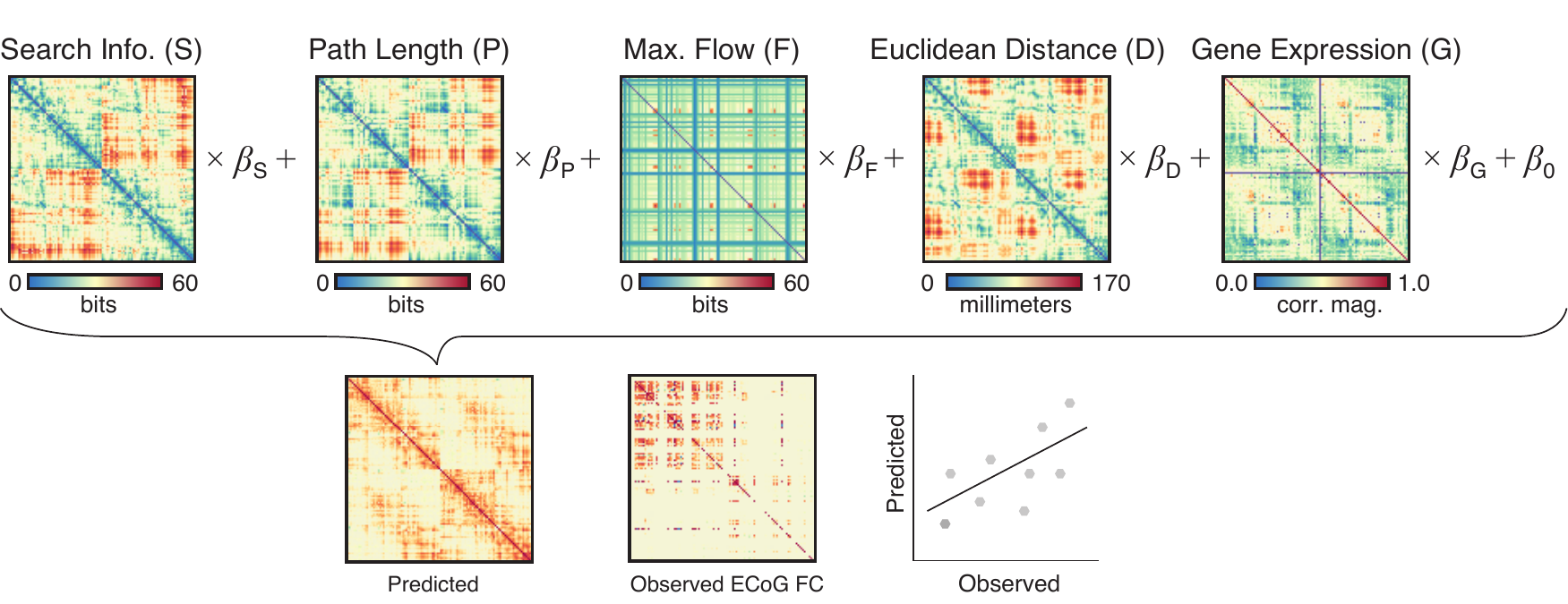}
\caption{\textbf{Schematic of nested multi-linear regression modeling approach to predicting ECoG functional connectivity from brain structure, geometry, and genetics.} Subsets of the five predictor matrices displayed above are used in a multi-linear regression model to predict observed ECoG functional connectivity. We show scaled representations of search information (a measure of the hiddenness of shortest paths between regions), path length (the true shortest path between regions), max flow (a measure of the total information flow between regions), Euclidean distance (the physical distance between regions), and gene coexpression (the correlation of gene expression profiles between regions). The correlation between the output of this trained model and the observed ECoG functional connectivity is used to measure its validity.\label{predictive_accuracy_methods}}
\end{figure}

We begin by assessing whether and how brain structure, geometry, and genetics influence functional connectivity during preictal and ictal periods. To this end, we investigated a series of nested multi-linear models (MLM) that generated predictions of functional connection weights (see \textbf{Figure \ref{predictive_accuracy_methods}}). For this purpose, we generated a list of five predictors as follows: \textbf{D} = $[\text{D}_{ij}]$, the Euclidean distance between region $i$ and region $j$; \textbf{G} = $[\text{G}_{ij}]$, the Pearson correlation between the gene expression profiles of region $i$ and region $j$; \textbf{S}=$[\text{S}_{ij}]$, the search information between region $i$ and region $j$, a measure of the ``hiddenness'' of the shortest anatomical path between the regions; \textbf{P}=$[\text{P}_{ij}]$, the length of the true shortest path length between region $i$ and region $j$; and \textbf{F}=$[\text{F}_{ij}]$, the maximum flow between region $i$ and region $j$, a measure that treats edge weights as capacities and sends units of flow from region $i$ to region $j$ while respecting maximum capacities. Search information, path length, and maximum flow all represent structural measures that capture different potential scales of communication dynamics that are supported by the network architecture. These three measures were computed from an undirected and weighted connectivity matrix, $A \in \mathbb{R} ^ {N \times N}$, whose edge weights were equal to the streamlines detected between regions normalized by the geometric mean of the regional volumes.\\

\begin{figure}[H]
\centering
\includegraphics[width=150mm]{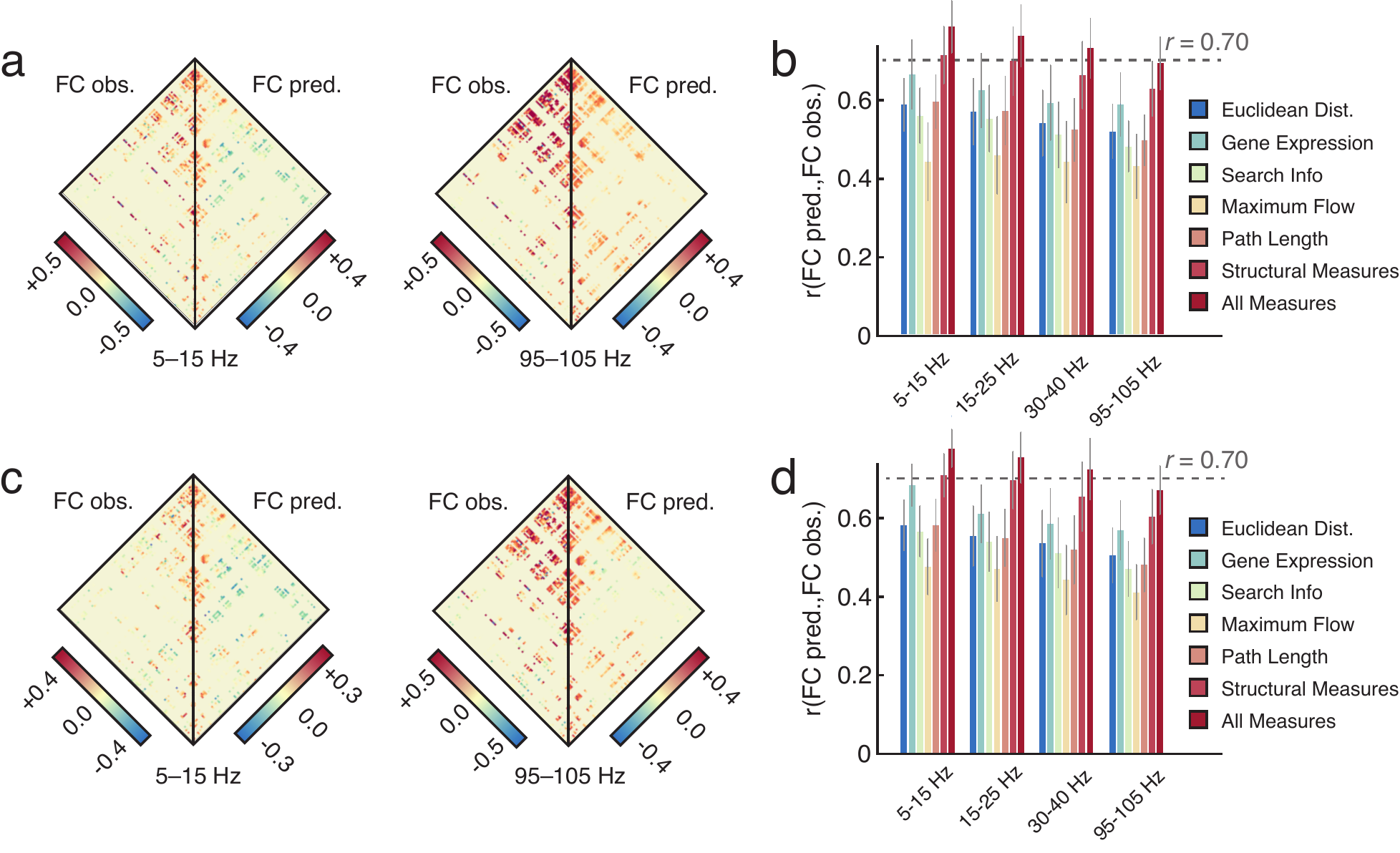}
\caption{\textbf{Predicting ECoG Functional Connectivity in Ictal and Preictal States.} We show the observed and best prediction of ECoG functional connectivity in the highest and lowest frequency bands for both preictal \textit{(a)} and ictal \textit{(c)} data. In panels \textit{(b)} and \textit{(d)} we show the performance of each single parameter model, as well as the performance of the model combining search information, maximum flow, and path length, which we denote with the phrase ``Structural Measures'' and which intuitively assesses the combined effectiveness of the three communication policies. Finally, we also show the performance of the full model, which combined all parameters. For all bar graphs, error bars indicate 95\% confidence intervals.\label{predictive_accuracy_results}}
\end{figure}

\noindent We constructed models with all combinations of predictors, from single factor models to models with all five predictors. All trained models performed remarkably well on both preictal and ictal functional connectivity. The correlation between predicted and observed functional connectivity was consistently greater than $r = 0.3734$ in both datasets. Moreover, all models with combinations of at least two predictors produced predicted functional connectivity matrices that were correlated with the observed functional connectivity matrices at an $r \geq 0.5182$ in ictal and preictal data (see \textbf{Figure \ref{predictive_accuracy_results}}). The final model, which contained all predictors, performed remarkably well, and produced predicted functional connectivity matrices that were correlated with the observed functional connectivity matrices with an $r \geq 0.6647$ across all frequency bands and across both ictal and preictal data.\\

\noindent Next we asked whether predictions were stronger or more accurate during the ictal states or during the preictal states. To address this question, we defined the model quality as the correlation between the true functional connectivity and the predicted functional connectivity. Notably, we observed that model quality was significantly different in preictal and ictal data ($p < 10^{-16}$ for all frequency bands using a paired $t$-test with Fisher's $rho$-to-$z$ transform). Specifically, preictal predictions outperformed ictal predictions in the three higher frequency bands ($r_pi - r_i > 0.0219$) and ictal predictions outperformed preictal predictions in the lowest frequency band ($r_i - r_pi = 0.0378$). This pattern of observations suggests that there are marked differences in the utility of the underlying models in explaining brain dynamics across states, perhaps reflecting the existence of distinct biological processes associated with seizure generation, propagation, and termination.

\subsection*{Preictal and ictal predictions generalize well}

\begin{figure}[H]
\centering
\includegraphics[width=160mm]{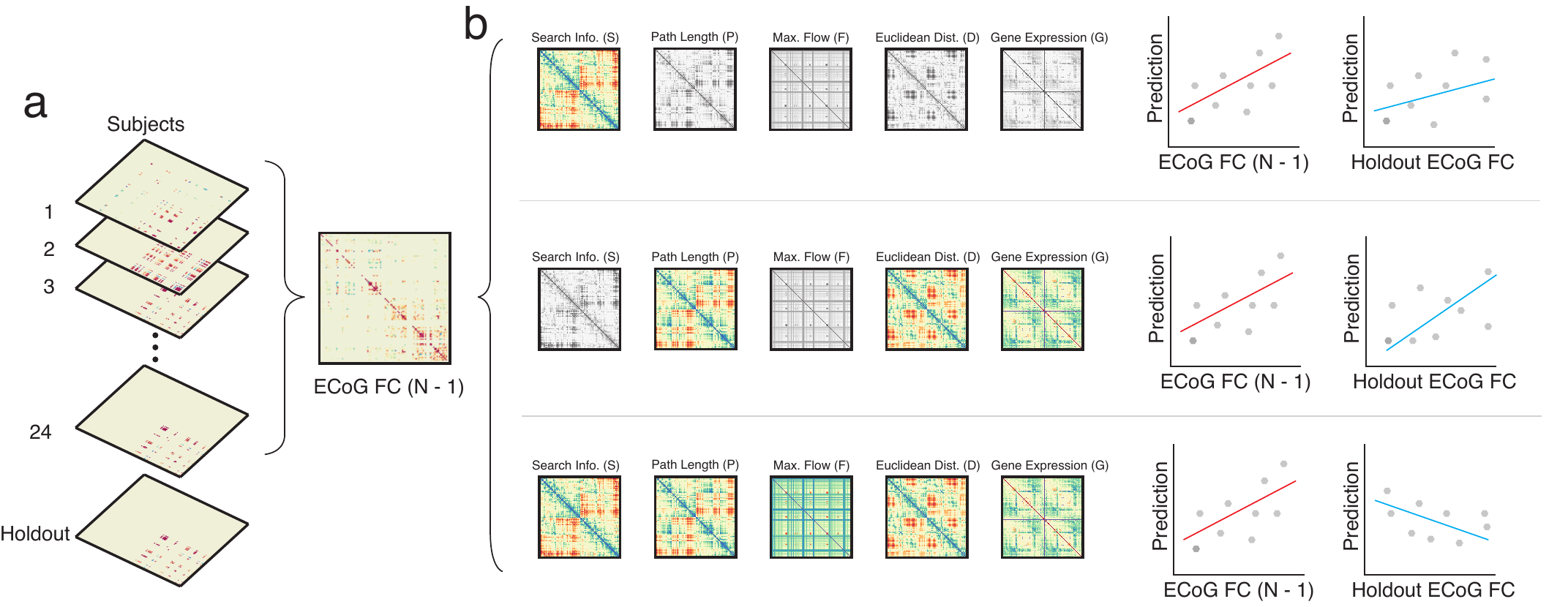}
\caption{\textbf{Cross-Validation to Assess Generalizability of Predictions.} \textit{(a)} Leaving out one subject, we generate a training functional connectivity matrix by averaging the functional connectivity matrices of the remaining subjects. \textit{(b)} We train separate multilinear regression models using each possible combination of predictors by maximizing the value of the Pearson correlation coefficient between their predicted functional connectivity matrix and the observed average functional connectivity matrix of the training set. We then evaluate the strength of these predictions by computing the Pearson correlation coefficient between this predicted functional connectivity matrix and the observed holdout functional connectivity matrix. \label{group_predictions_methods}}
\end{figure}

Thus far, we have shown that ictal and preictal functional connectivity can be partially explained by combinations of genetic, physical, and structural network features. We next turn to the question of whether predictions that are trained on one subset of subjects are also generalizable to other subsets of subjects. This question is important because such generalizability would be reflective of reliability and robustness of the architectural substrates for functional ECoG dynamics. To address this question, we chose to employ a standard cross-validation protocol in which one subject is ``held-out'' and a model is trained on the remaining subjects, and then the same model is tested on the single hold-out (see \textbf{Figure \ref{group_predictions_methods}}). \\

\begin{figure}[H]
\centering
\includegraphics[width=110mm]{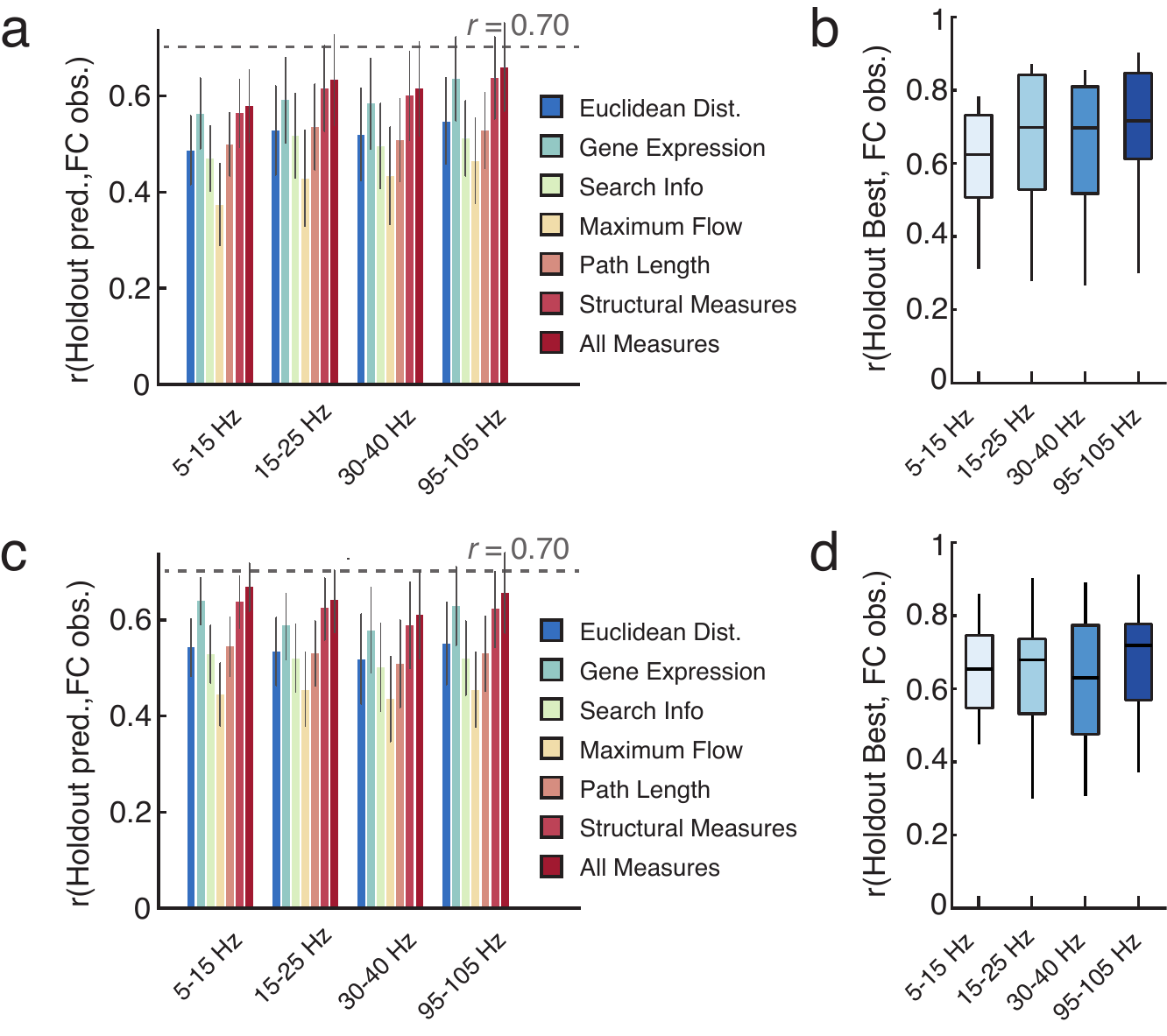}
\caption{\textbf{Predictions of ECoG Functional Connectivity in Ictal and Preictal States Using Leave-One-Out Cross-Validation.} Across frequency bands, we show the performance of each single parameter model, as well as the performance of the model combining search information, maximum flow, and path length, which we denote with the phrase ``Structural Measures'' and which intuitively assesses the combined effectiveness of the three communication policies. Finally, we also show the performance of the full model, which combined all parameters. Performance results are separately displayed for preictal (panel \textit{(a)}) and ictal (panel \textit{(c)}) data. For all bar graphs, error bars indicate 95\% confidence intervals. We also show the performance of the optimal set of training parameters in each frequency band in both preictal (panel \textit{(b)}) and ictal (panel \textit{(d)}) data. Performance is measured by average predictive accuracy in the holdout model of cross-validation shown in \textbf{Figure \ref{group_predictions_methods}}. \label{group_predictions_results}}
\end{figure}

\noindent Using this leave-one-out cross-validation approach, we find that even single predictor models provide surprisingly accurate predictions of preictal and ictal functional connectivity matrices in this group-training setting. On average, all single predictor models produced predicted functional connectivity matrices that were correlated with the observed functional connectivity matrices at an $r \geq 0.3721$ on preictal data in all frequency bands, and at an $r \geq 0.4180$ on ictal data in all frequency bands (see \textbf{Figure \ref{group_predictions_results}}). Interestingly, the model that performed the best in this setting was in fact the same for all frequency bands and consisted of gene coexpression, maximum flow, and shortest path length, omitting the Euclidean distance and the search information. These results suggest that network geometry and the hiddenness of shortest paths in the network are less important for generalizabile predictions than markers of the length of the true shortest path between regions, the total information flow between regions, and the correlation of gene expression profiles between regions.\\

\noindent Next we asked whether the quality of predictions based on the leave-one-out cross-validation was stronger or more accurate during the ictal states or during the preictal states. To address this question, we used the same definition of model quality: the correlation between the true functional connectivity and the predicted functional connectivity. Across all models, we find that ictal predictions significantly outperform preictal predictions in the lowest frequency band ($r_i - r_{pi} = 0.0563$, $p = 2.77 \times 10^{-20}$ using a paired $t$-test with Fisher's $\rho$-to-$z$ transform) while preictal predictions significantly outperform ictal predictions in all other frequency bands ($r_i - r_{pi} < -0.0136$, $p < 1.47 \times 10^{-09}$ using a paired $t$-test with Fisher's $\rho$-to-$z$ transform). These results suggest that measures of brain structure, geometry, and genetics can accurately predict ECoG functional connectivity in both ictal and preictal states, with ictal dynamics being more easily explained at low frequencies and preictal dynamics being more easily explained at midrange and high frequencies.\\

\subsection*{Single subject preictal and ictal predictions generalize well}

\begin{figure}[H]
\centering
\includegraphics[width=160mm]{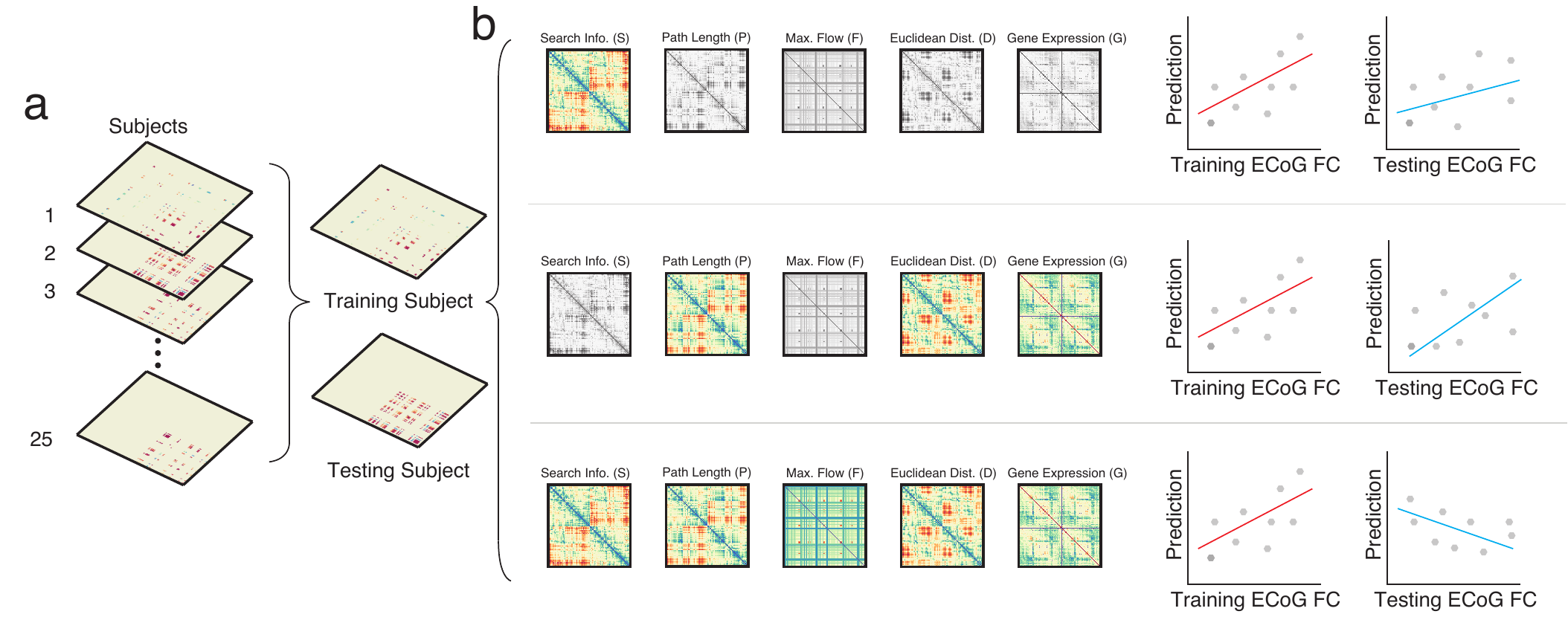}
\caption{\textbf{Single-Subject Prediction Approach.} \textit{(a)} In this approach, we first choose a single subject to represent the testing set, and then we consider all other single subjects as training datasets. \textit{(b)} We train separate multilinear regression models using each possible combination of predictors by maximizing the Pearson correlation coefficient between their predicted functional connectivity matrix and the observed average functional connectivity matrix of the training set. We then evaluate the strength of these predictions by computing the Pearson correlation coefficient between this predicted functional connectivity matrix and the observed holdout functional connectivity matrix, averaged across all training datasets. \label{subject_predictions_methods}}
\end{figure}

A critical and pervasive question in clinical applications is whether and to what degree individual differences in brain architecture and dynamics determine the choice of treatment for a given patient, as well as their response to that treatment. The answer to this question depends in large part on the degree to which the architecture and function of the neural system is conserved or variable across individuals. The degree to which a model is dependent on individual heterogeneity can be assessed to some degree using leave-one-out cross-validation, as we discussed and implemented in the previous section. However, an even more stringent method to assess individual heterogeneity is to ask whether a model built from a single subject can be used to predict the neurophysiological dynamics of a different subject. To explicitly assess such fine-scale pairwise heterogeneity, we used the functional connectivity matrix of a single subject to predict the functional connectivity matrix of a different subject, and we quantified the quality of those predictions (see \textbf{Figure \ref{subject_predictions_methods}}).\\

\noindent As in the group training setting, we asked whether the quality of predictions based on the single-subject prediction approach were stronger or more accurate during the ictal states or during the preictal states. To address this question, we used the same definition of model quality: the correlation between the true functional connectivity and the predicted functional connectivity. Across all models of single subjects, we find that ictal predictions significantly outperform preictal predictions in the lowest frequency band ($r_i - r_{pi} = 0.0752$, $p = 1.380 \times 10^{-27}$ using a paired $t$-test with Fisher's $\rho$-to-$z$ transform) and also in the second lowest frequency band ($r_i - r_{pi} = 0.0128$, $p = 2.276 \times 10^{-4}$ using a paired $t$-test with Fisher's $\rho$-to-$z$ transform). In contrast, we find that preictal predictions significantly outperform ictal predictions in both of the higher frequency bands ($r_i - r_{pi} < -0.012$, $p < 4.075 \times 10^{-8}$ using a paired $t$-test with Fisher's $\rho$-to-$z$ transform). Again, these results suggest that measures of brain structure, geometry, and genetics can accurately predict ECoG functional connectivity in both ictal and preictal states, with ictal dynamics being more easily explained at low frequencies and preictal dynamics being more easily explained at high frequencies.\\

\begin{figure}[H]
\centering
\includegraphics[width=160mm]{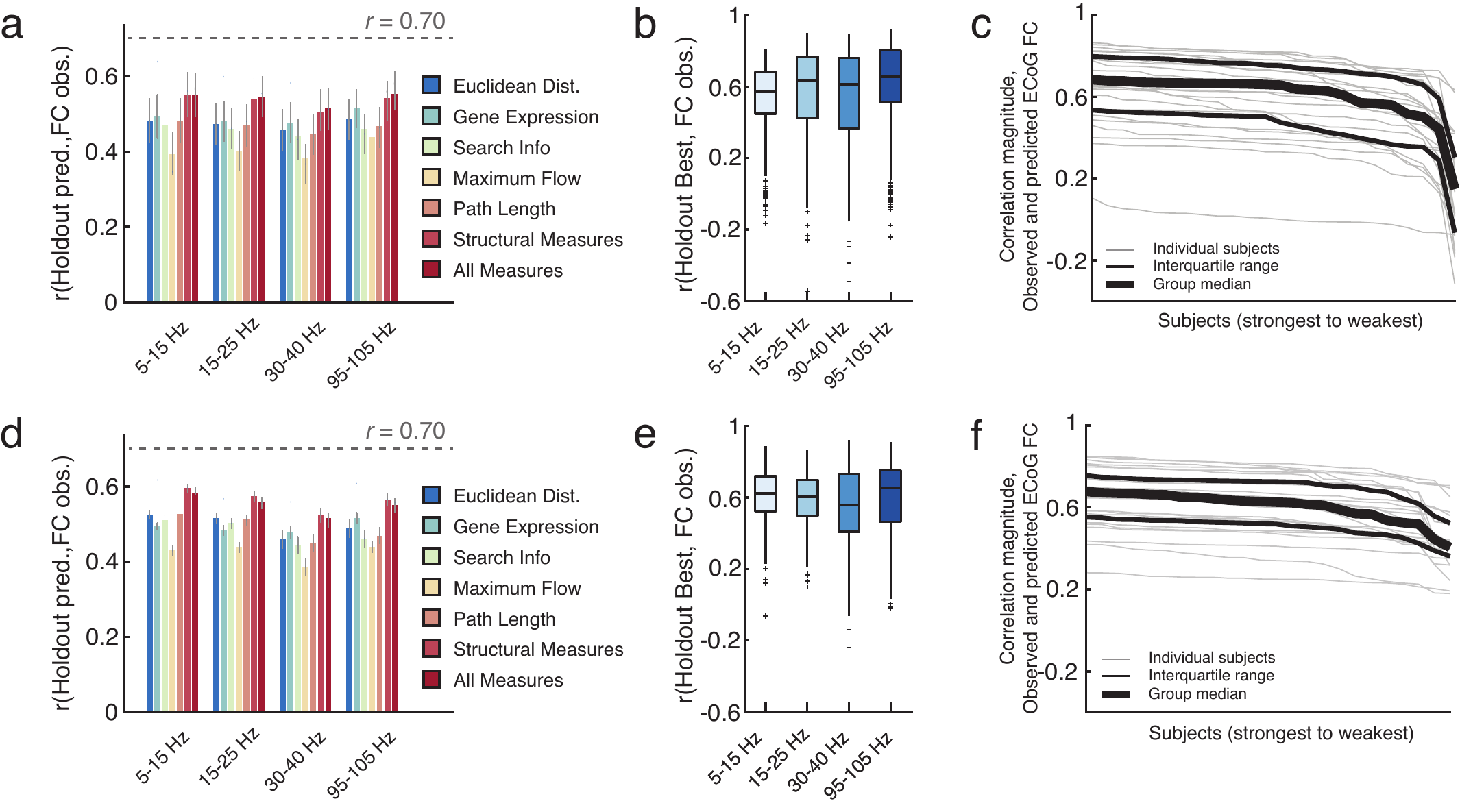}
\caption{\textbf{Predictions of ECoG Functional Connectivity in Ictal and Preictal States Using the Single-Subject Approach.} Across frequency bands, we show the performance of each single parameter model, as well as the performance of the model combining search information, maximum flow, and path length, which we denote with the phrase ``Structural Measures'' and which intuitively assesses the combined effectiveness of the three communication policies. We also show the performance of the full model, which combined all parameters. Performance results are separately displayed for preictal (panel \textit{(a)}) and ictal (panel \textit{(d)}) data. For all bar graphs, error bars indicate 95\% confidence intervals. We also show the performance of the optimal set of training parameters in each frequency band in both preictal (panel \textit{(b)}) and ictal (panel \textit{(e)}) states. Performance is measured by average predictive accuracy in the single subject model shown in \textbf{Figure \ref{subject_predictions_methods}}. Finally, we show the distribution of the quality of predictions made for each subject, where each line represents the performance of predictions of a single subject's functional connectivity ordered from best to worst in both preictal (panel \textit{(c)}) and ictal (panel \textit{(f)}) states. \label{subject_predictions_results}}
\end{figure}

\noindent Next, we turned to the important question of whether model performance differed significantly across different subjects. Even if the general pairwise predictive accuracy was quite good, it remains possible that one (or a few) subject(s) provided significantly poorer or significantly better predictions for other subjects. To assess such individual variability, we considered each subject separately, and then we trained a model on the functional connectivity of every other subject. Next, we ordered subjects by the value of the correlation coefficient between the predicted functional connectivity matrix and the true functional connectivity matrix (Fig.~\ref{subject_predictions_results}c,f). We found that the worst preictal predictions are significantly worse than the worst ictal predictions, and in fact, even the worst ictal predictions do reasonably well ($r_i - r_{pi} > 0.311$, $p < 8.3 \times 10^{-8}$ using a paired $t$-test for the  ten worst predictions of each subject's functional connectivity). These results provide some initial support for the notion that ictal data contains features of network dynamics that are more conserved across subjects than features of network dynamics in preictal data. While seizure patterns can be heterogeneous across different states, non-seizure states must encompass a much larger potential range of actions and, therefore, are less likely to be predicted by the multi-linear model used here.

\subsection*{Beta values distinguish between preictal and ictal predictions}

In the previous sections, we presented results on the performance of various models in predicting ECoG functional connectivity in preictal and ictal states. Now we turn to the important question of whether individual features in the models contribute equally to the ictal and preictal predictions. If mechanisms are conserved then all features should contribute similarly across ictal and preictal states, whereas if mechanisms differ then all features should contribute differentially to ictal and preictal states. We hypothesized that mechanisms would differ. Specifically, based on recent studies of the relation between white matter and seizure duration or severity \citep{chiang2016white,labate2015white}, we hypothesized that ictal functional connectivity would rely more heavily on the brain's structural network architecture and communication policies atop that architecture than preictal functional connectivity.\\

\noindent To test this hypothesis, we train a model with every predictor matrix separately on each frequency band of the preictal and ictal data, and we obtain $\beta$ weights placed on each predictor matrix. This framework allowed us to isolate the redundant, synergistic, and unique contributions made by individual factors, and as they appeared in models either in isolation or together with other factors. To show robustness of these results to variability in the underlying data, we train separate models while omitting one subject in each model. We then compute the average normalized difference between $\beta$ weights assigned to each parameter in the model for ictal and preictal datasets with all parameters present. \\

\noindent When training on ictal data compared to preictal data, we find significantly lower $\beta$ weights associated with Euclidean distance ($\beta_{ictal} - \beta_{preictal} < -0.063$, $p<5.8242 \times 10^{-13}$ for all frequency bands using a paired $t$-test). We also find significantly lower $\beta$ weights associated with gene coexpression ($\beta_{ictal} - \beta_{preictal} < -0.1136$, $p < 2.204 \times 10^{-19}$ for the two highest frequency bands using a paired $t$-test) when training on ictal data compared to preictal data. Simultaneously, we found significantly higher $\beta$ weights associated with maximum flow ($\beta_{ictal} - \beta_{preictal} > 0.0902$, $p < 3.080 \times 10^{-16}$ for the two highest frequency bands using a paired $t$-test) when training on ictal data compared to preictal data. Consistent with our hypothesis, these data demonstrate that ictal functional connectivity relies more heavily on the brain's structural network architecture (and communication policies putatively enacted upon it) than preictal functional connectivity. Moreover, the data demonstrate that preictal functional connectivity depends more heavily on conserved patterns of gene coexpression than ictal functional connectivity. \\

\begin{figure}[H]
\centering
\includegraphics[width=150mm]{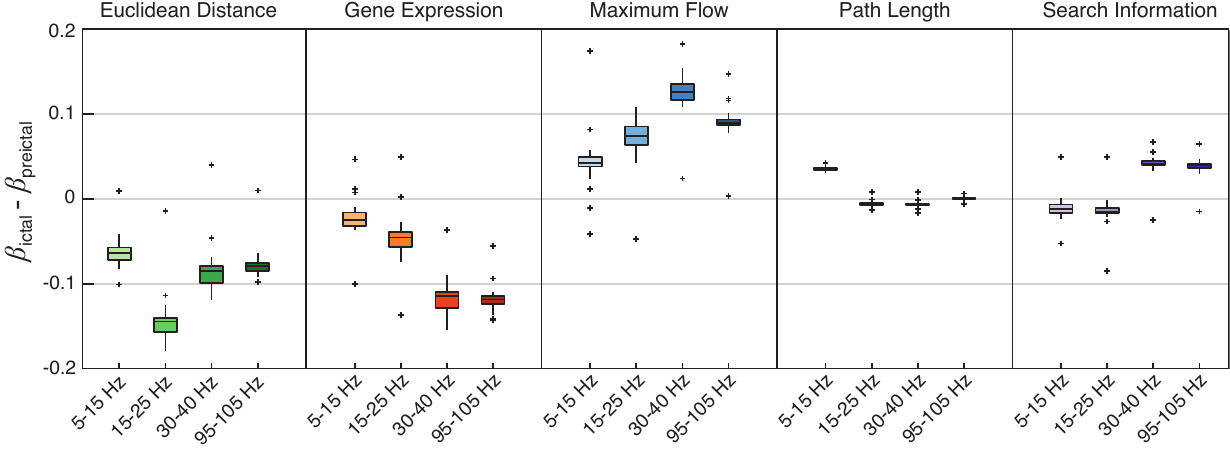}
\caption{\textbf{Ictal and Preictal Dynamics Differentially Depend on Brain Structure, Geometry, and Genetics.} In the four frequency bands of interest, we show the difference between the $\beta$ weights of features in models of ictal data and the $\beta$ weights of features in models of preictal data. From left to right, we show these $\beta$ weight differences for the following features: Euclidean distance, gene coexpression, maximum flow, path length, and search information. Boxplots show median (black line), first to third quartile of data (box), range excluding outliers (whiskers), and outliers (`+'). Color saturation marks relative frequency, with less saturation for lower frequencies and greater saturation for higher frequencies. \label{beta_values_results}}
\end{figure}

\section*{Discussion}

In this work, we trained multi-linear regression models to predict the architecture of functional brain networks constructed from preictal and ictal epochs extracted from electrophysiological (ECoG) recordings in 25 patients with medication refractory epilepsy. We hypothesized that the performance of these models would differ in functional networks constructed from ictal epochs and functional networks constructed from preictal epochs. By quantifying the strength of the model using the correlation coefficient between the true functional connectivity and the predicted functional connectivity, we find that model strength is significantly different in the ictal and preictal data, but that the most powerful models in both ictal and preictal data reach the same level of prediction accuracy. Critically, we find notable differences in the relative weight placed on structural, geometric, and genetic features between ictal and preictal data. Significantly higher weight on structural features in the ictal data suggests that the functional brain network dynamics manifest in seizures rely on -- and may directly propagate along -- underlying white matter structure that is largely conserved across humans, in a sterotyped manner consistent with specific putative communication policies. Significantly higher weight on Euclidean distance and gene coexpression in the preictal data suggests that the more general functional brain network dynamics occurring outside of the seizure state rely on conserved principles of geometry and shared genetic constraints. Collectively, our findings fundamentally extend our understanding of how seizure dynamics depend upon common conserved anatomical organization while breaking with constraints of local geometry and instrinsic genetic markers of neurophysiological function.

\subsection*{ECoG functional connectivity is partially explained by brain structure, geometry, and genetics}

Across both ictal and preictal states, even single parameter models perform remarkably well in predicting ECoG functional connectivity, and in fact, also generalize well from a single sample, here reflecting a single subject. The consistently high prediction accuracy across models is particularly notable given that the features are not specific to the subjects being tested. Indeed, gene coexpression data is taken from the Allen Brain Institute, and represents an agglomeration over a set of healthy adult donors \citep{jones2009allen, hawrylycz2012anatomically, sunkin2013allen}. Similarly, the structural features that we study -- and that reflect various communication policies -- are computed from the average structural connectivity of a healthy adult population \citep{betzel2016optimally, betzel2017modular, betzel2017diversity}. The fact that ECoG functional connectivity is so well-predicted by models derived from these features suggests that there exist highly conserved principles of human brain anatomy, geometry, and genetics that form pervasive constraints on neurophysiological dynamics in disease across both persistent baseline and transient pathological states. It would be interesting in future work to examine the degree to which these models could potentially be enhanced with subject-level structural data, and to determine whether the same models could be used to predict different phases of seizure initiation, propagation, and termination \citep{khambhati2015dynamic,burns2014network}.

\subsection*{The role of white matter architecture in ictal network dynamics}

We found that structural connectivity features were weighted far greater in models of the seizure state than in models of the preictal state. Interestingly, this observation suggests that the underlying anatomy of white matter structure plays an enhanced role in governing the behavior of neurophysiological dynamics during seizures. Notably, we found the greatest difference in the weights of the maximum flow feature, which was the most topologically global of the structural measures that we studied. Commonly used in information theory \citep{ahlswede2000network}, maximum flow measures the total amount of some item that can be sent along the network from one point to another by treating the strength of edges as capacities for that edge \citep{ford1956maximal}. In the context of neurophysiological processes, this feature could be considered a medium for the evolution and consequent propagation of seizure dynamics through the network \citep{khambhati2015dynamic,burns2014network}.\\

\noindent More generally, we found that structural connectivity features varied in terms of their ability to accurately predict functional connectivity, with path length and search information performing slightly better than max flow. Along with metrics such as navigability \cite{seguin2018navigation}, diffusion \cite{abdelnour2014network}, and communicability \cite{estrada2008communicability}, these measures, and the assumptions about neural communication that they entail \cite{avena2018communication}, have all been used in past studies to relate structural and functional connectivity (usually estimated at rest), \emph{via} a particular communication strategy or policy \cite{goni2014resting, messe2015predicting, messe2015relating}. Typically, measures that are strongly associated with functional connectivity are also treated as being more likely to represent the true underlying process by which that functional connectivity pattern was generated. Here, however, in addition to comparing overall fit, we compare a subset of these measures and their sensitivities to seizure dynamics, identifying max flow as the measure with the greatest sensitivity, suggesting that it may be of greater relevance. In the future, explicitly causal and mechanistic models could be used to more deliberately test this hypothesis \cite{frassle2018generative}.\\

\noindent In comparing ictal and preictal regression weights, the emphasis on structural connectivity is coupled with a relative disregard for the constraints of gene coexpression as well as the physical distance between regions. This latter independence of dynamics from geometry is striking, particularly in light of the fact that physical distance between brain regions seems to be a pervasive constraint in healthy functional connectivity \citep{stiso2018spatial}. Indeed, much theoretical work has sought to frame the dependence of connectivity on distance in light of a trade-off between communication cost and communication efficiency \citep{bullmore2012economy}. Interestingly, our results suggest that functional connectivity during ictal periods is more strongly predicted by the underlying topological pattern of white-matter connections than by any signal conduction effects through the physical volume of the brain \citep{chiang2016white,labate2015white}. This insight suggests that functional connections observed during a seizure may actually reflect brain dynamics in more distant regions than commonly expected, underscoring the increasing need to understand distributed markers of pathology \citep{proix2017individual, jirsa2017virtual}.

\subsection*{Clinical applicability}

In addition to offering insights into the generic architectural principles of the human brain that impact seizure dynamics, our methodology could be used to extend our understanding, models, and predictions of subject-level pathology and response to intervention in future applications. By combining our general computational framework with the goals of virtual cortical resection \citep{khambhati2016virtual}, it could be useful to test how the resection of specific cortical or subcortical volumes affects various structural measures in a manner that might be predicted to hamper seizure spread by decrementing the correspondence between ictal dynamics and white matter architecture. Furthermore, a marked limitation of current IEEG studies, both for epilepsy surgery and for basic cognitive neuroscience, is the fact that ECoG measurements provide only partial brain coverage, thereby yielding incomplete representations of epileptic networks, in some cases even missing the seizure onset zone itself or other regions of seizure spread \citep{khambhati2015dynamic}. It would be interesting to use our model to predict functional connectivity in areas of the brain that are not covered by stereo or grid electrodes, or perhaps to inform which areas of the brain absolutely require coverage in order for us to obtain accurate assessments of the pathology or predictions of response to resection or ablation. Finally, our model could be used to determine which time periods throughout the seizure or in preictal states are most relevant to the pathology of the individual, rather than to pathology shared across patients with focal epilepsy. Given the differential predictions of generic structural, geometric, and genetic data for ictal vs. preictal dynamics in low vs. high frequency bands, our results suggest that individual-specific information might be most prevalent in high frequency ictal dynamics and low frequency preictal dynamics. In future work, this suggestion could be directly tested and evaluated for utility in clinical contexts.

\subsection*{Methodological considerations}

Composite structural connectivity was used to compute the structural parameters for each model, meaning that none of the individual subject data has been incorporated into the training of this model. Replacing average connectivity with subject-specific connectivity might allow trained models to better predict functional connectivity of a single subject, and thus perhaps identify drivers of the epileptic network. Additionally, subjects included in this study had seizures originating in different regions of the brain, which may worsen the quality of the predictions made. To realize potential clinical uses, future work will require incorporating further unique subject information, including patient specific imaging data, to improve the specificity of results.\\

\noindent A second limitation concerns the gene coexpression data. These data were obtained from the Allen Brain Institute Human Brain Atlas \cite{hawrylycz2012anatomically} and represent the average coexpression pattern of only two individual subjects that differed in age and demographics. It remains unclear whether the coexpression patterns obtained from this small but unique dataset are, in fact, representative of typical individuals.\\

\noindent A third limitation concerns the reconstruction of white matter fiber networks from diffusion MRI data using tractography. This reconstruction procedure is prone to systematic errors \cite{maier2016tractography,reveley2015superficial,thomas2014anatomical}, which may introduce bias into any measure computed from a white matter network, including the measures of search information, max flow, and path length studied here. Although these biases remain potential confounds, advances in acquisition hardware, processing strategies \cite{yeh2018population}, and tractography algorithms \cite{pestilli2014evaluation} will help mitigate these issues in future work.

\section*{Conclusion}

Medically refractory epilepy remains a critical burden to our society, and to the patients that suffer from it. Novel technologies including enhacements to resection, laser ablation, and neurostimulation, afford increasing specificity for treatment but critically depend on the accurate fusion of multimodal neuroimaging data to map anatomical targets driving seizure dynamics. Here we propose a parsimonious model that explains how the large-scale anatomical networks and shared genetic constraints shape inter-regional communication in focal epilepsy. The work offers broad insights into the generic architectural principles of the human brain that impact seizure dynamics, and could be extended to further our understanding, models, and predictions of subject-level pathology and response to intervention.

\section*{Acknowledgments}
We thank Jennifer Stiso, Xiaosong He, and Richard Rosch for helpful comments on earlier versions of this manuscript. The work was supported by a grant to B.L. and D.S.B from the National Institute of Neurological Disorders and Stroke (R01 NS099348). D.S.B., P.G.R., and R.F.B. also acknowledge support from the John D. and Catherine T. MacArthur Foundation, the Alfred P. Sloan Foundation, the ISI Foundation, the Paul Allen Foundation, the Army Research Laboratory (W911NF-10-2-0022), the Army Research Office (Bassett-W911NF-14-1-0679, Grafton-W911NF-16-1-0474, DCIST- W911NF-17-2-0181), the Office of Naval Research, the National Institute of Mental Health (2-R01-DC-009209-11, R01 – MH112847, R01-MH107235, R21-M MH-106799), the National Institute of Child Health and Human Development (1R01HD086888-01), and the National Science Foundation (BCS-1441502, BCS-1430087, NSF PHY-1554488 and BCS-1631550). The content is solely the responsibility of the authors and does not necessarily represent the official views of any of the funding agencies.

\section*{Conflict of Interest}

The authors declare that they have no conflicts of interest related to this work.

\section*{Materials and Methods}

\subsection*{ECoG data collection and preprocessing}

\textit{Ethics Statement.} All patients at the Hospital of the University of Pennsylvania included in this study gave written informed consent in accordance with the Institutional Review Board of the University of Pennsylvania. \\

\noindent \textit{Electrophysiology recordings.} Twenty five patients, twenty at the University of Pennsylvania and five at the Mayo Clinic, undergoing surgical treatment for medically refractory epilepsy underwent implantation of subdural and depth electrodes to localize the seizure onset zone after presurgical evaluation with scalp EEG recording of ictal epochs, MRI, PET and neuropsychological testing suggested that focal resection may be a therapeutic option. Patients were then deemed candidates for implantation of intracranial electrodes to better define the epileptic network. De-identified patient data was retrieved from the online International Epilepsy Electrophysiology Portal (IEEG Portal) \citep{kini2016data}.\\

\noindent For patients at the University of Pennsylvania, ECoG signals were recorded and digitized at 500 Hz sampling rate using Nicolet C64 amplifiers and pre-processed to eliminate line noise. Cortical surface electrode configurations (Ad Tech Medical Instruments, Racine, WI), determined by a multidisciplinary team of neurologists and neurosurgeons, consisted of linear and two-dimensional arrays (2.3 mm diameter with 10 mm inter-contact spacing) and linear depths (1.1 mm diameter with 10 mm inter-contact spacing). Signals were recorded using a referential montage with the reference electrode being chosen by the clinical team to be distant to the site of seizure onset. Recording spanned the duration of a patient's stay in the epilepsy monitoring unit. We note that some recording information was unavailable for patients treated at the Mayo Clinic.\\

\noindent \textit{Epileptic events.} From patients at the University of Pennsylvania, we analyzed 41 partial seizures (simple and complex) and 57 partial seizures that generalized to surrounding tissue. Of the 20 Hospital of the University of Pennsylvania epilepsy patients in the study cohort, 8 patients exhibited strictly complex-partial seizures that secondarily generalized (distributed events), 6 patients exhibited strictly simple-partial or complex-partial seizures that did not secondarily generalize (focal events), and 6 patients exhibited a combination of distributed events and focal events. Seizure type, onset time, and onset localization were marked as a part of routine clinical workup. Patient demographic data for these patients is included in Table \ref{HUPTable}.\\

\begin{table}[h!]
\centering
\begin{tabular}{ | c | c | c | c | c | }
\hline
	Patient ID & Age & Gender & Resected Region & Seizure Type \\ \hline
	HUP064 & 21 & M & LFL & CPS+GTC \\ \hline
	HUP065 & 37 & M & RTL & 3 CPS+GTC \\ \hline
	HUP068 & 28 & F & RTL & 4 CPS+GTC, 1 CPS \\ \hline
	HUP070 & 33 & M & LFPL & 8 SPS \\ \hline
	HUP073 & 40 & M & RFL & 5 CPS+GTC \\ \hline
	HUP074 & 25 & F & LTL & 1 CPS+GTC, 4 CPS \\ \hline
	HUP082 & 56 & F & RTL & 6 CPS \\ \hline
	HUP086 & 25 & F & LTL & 2 CPS+GTC \\ \hline
	HUP087 & 24 & M & LFL & 2 CPS \\ \hline
	HUP088 & 35 & F & LTL & 1 CPS+GTC, 2 CPS \\ \hline
	HUP094 & 48 & F & RTL & 3 CPS+GTC \\ \hline
	HUP105 & 39 & M & RTL & 2 CPS+GTC \\ \hline
	HUP106 & 45 & F & LTL & 1 Aura, 3 CPS+GTC, 1 CPS \\ \hline
	HUP107 & 36 & M & RTL & 1 Aura, 1 CPS+GTC, 5 CPS \\ \hline
	HUP111 & 40 & F & RTL & 7 SPS, 12 CPS \\ \hline
	HUP113 & 47 & M & LTL & 1 CPS+GTC \\ \hline
	HUP075 & 57 & F & LTL & CPS \\ \hline
	HUP078 & 54 & M & LTL & 5 CPS \\ \hline
	HUP080 & 41 & F & LTL & 4 CPS+GTC \\ \hline
	HUP083 & 29 & M & LPL & 9 CPS, 3 SPS, 1 CPS+GTC \\ \hline
\end{tabular}
\caption{\textbf{Demographic Information for Patients Treated at the Hospital of the University of Pennsylvania.} Patient datasets accessed through IEEG Portal (https://www.ieeg.org). Age at Phase II monitoring. Gender: M, male; F, female. Seizure types are CPS (complex partial seizure), SPS (simple partial seizure), CPS+GTC (complex partial with secondary generalization). Counted seizures were recorded in the epilepsy-monitoring unit. Resected regions are LFL (left frontal lobe), RFL (right frontal lobe), LTL (left temporal lobe), RTL (right temporal lobe), LFPL (left frontoparietal lobe), and RFPL (right frontoparietal lobe). \label{HUPTable}}
\end{table}

\noindent In the IEEG Portal, information regarding the type and number of seizures is unavailable for patients treated at the Mayo Clinic. For these patients, we therefore present a limited set of demographic data; see Table \ref{MayoTable}.\\

\begin{table}[h!]
\centering
\begin{tabular}{ | c | c | c | c | }
\hline
	Patient ID & Gender & Resected Region \\ \hline
    Study012 & M & RFL \\ \hline
	Study016 & F & RFTL \\ \hline
	Study017 & M & RTL \\ \hline
	Study022 & F & LTL \\ \hline
	Study028 & M & LFPL \\ \hline
\end{tabular}
\caption{\textbf{Demographic Information for Patients Treated at the Mayo Clinic.} Patient datasets accessed through IEEG Portal (https://www.ieeg.org). Gender: M, male; F, female. Resected regions are LFL (left frontal lobe), RFL (right frontal lobe), LTL (left temporal lobe), RTL (right temporal lobe), LFPL (left frontoparietal lobe), and RFPL (right frontoparietal lobe). \label{MayoTable}}
\end{table}

\noindent \textit{Clinical marking of the seizure onset zone.} The seizure state was deemed to span between the clinically-marked earliest electrogaphic change (EEC) \citep{litt2001epileptic} and seizure termination. The pre-seizure state was the period immediately preceding the seizure state and of equal duration. We refer to the pair of pre-seizure and corresponding seizure states as an \textit{event}. For patients at the University of Pennsylvania, the seizure onset zone was marked on the Intracranial EEG (IEEG) according to standard clinical protocol in the Penn Epilepsy Center. Initial clinical markings are made on the IEEG the day of each seizure by the attending physician: a board certified, staff epileptologist responsible for that inpatient's care. Each week these IEEG markings are vetted in detail, and then finalized at surgical conference according to a consensus marking of 4 board certified epileptologists. These markings on the IEEG are then related to other multi-modality testing, such as brain MRI, PET scan \citep{waxman2009society}, neuropsychological testing, and ictal SPECT scanning to finalize surgical approach and planning. This process is the standard of clinical care stipulated by the National Association of Epilepsy Centers (NAEC), and upheld at all certified Level-4 epilepsy centers in the United States. Information regarding seizure onset marking is unavailable for patients treated at the Mayo Clinic.\\

\noindent \textit{ECoG pre-processing.} Artifactual channels were discarded and the remaining channels were referenced to the average signal, pre-whitened by retaining the residuals after fitting a first-order autoregressive model to the referenced time series, stop-filtered to remove line noise and its harmonics, and bandpass filtered into canonical frequency bands associated with seizure activity: $\alpha - \theta$ (5-15 Hz); $\beta$ (15-25 Hz); low-$\gamma$ (30-40 Hz); high-$\gamma$ (95-105 Hz). For each subject and for each trial, we computed the inter-electrode functional connectivity as a zero-lag Pearson correlation coefficient averaged across trials. \\

\noindent \textit{Mapping electrode locations.} Electrode locations were identified via thresholding of each patient's post-implant CT image. Next, each patient’s CT and T1-weighted MRI images were co-registered via 3D rigid affine registration. These T1-weighted MRI images were then aligned to the standard MNI brain using diffeomorphic registration with the symmetric normalization (SyN) method \citep{avants2008symmetric}. The resulting transformations were used to warp the coordinates of the electrode centroids into MNI space. These locations were subsequently mapped to the MNI standard coordinate system using the FSL function \texttt{img2stdcoord}. We compared each electrode's location in MNI space to points (vertices) on the \textit{fsaverage} pial surface, and assigned each vertex to an electrode if the Euclidean distance between the two was less than 5 mm. Each surface vertex was also assigned to one of $N = 114$ cortical regions as defined by the AAL atlas \citep{tzourio2002automated}, thereby making it possible to map electrodes to brain regions. \\

\noindent \textit{Group-aggregated ECoG functional connectivity.} For every pair of brain regions, $i$ and $j$, and for each subject independently, we identified all electrode pairs, $u$ and $v$ where electrode $u$ was assigned to region $i$ and electrode $v$ was assigned to region $j$. We estimated their average connection weights to generate a subject-specific inter-regional ECoG functional connectivity matrix. We then estimated the connection weight $A_{ij}$ in the group-aggregated ECoG matrix by averaging connection weights over all subjects. We repeated this procedure separately for each of the four frequency bands, resulting in band-limited, whole-brain, inter-regional ECoG functional connectivity matrices.

\subsection*{Diffusion imaging data collection and preprocessing}

\noindent To study conserved structural features that may constrain ECoG network dynamics, we analyzed a group-representative, whole-brain structural connectivity network following \citep{betzel2017inter}. Specifically, we built a group-representative connectome by combining single-subject data from a cohort of 30 healthy adult participants. Each participant's structural network was reconstructed from diffusion spectrum images (DSI) in conjunction with state-of-the-art tractography algorithms to estimate the location and strength of large-scale interregional white matter pathways. Study procedures were approved by the Institutional Review Board of the University of Pennsylvania, and all participants provided informed consent in writing. Details of the acquisition and reconstruction have been described elsewhere \cite{betzel2016optimally, betzel2017modular, betzel2017diversity}. We studied a division of the brain into $N=114$ cortical regions \cite{cammoun2012mapping}. Based on this division, we constructed for each individual an undirected and weighted connectivity matrix, $A \in \mathbb{R}^{N \times N}$, whose edge weights were equal to the number of streamlines detected between region $i$ and region $j$, normalized by the geometric mean of region $i$'s and region $j$'s volumes: $A_{ij} = \frac{S_{ij}}{\sqrt{(V_i V_j)}}$.\\

\noindent The resulting network was undirected (i.e., $A_{ij} = A_{ji}$). These individual-level networks were then aggregated to form a group-representative network. This aggregation procedure can be viewed as a distance-dependent consistency thresholding of connectome data and the details have been described elsewhere \cite{mivsic2015cooperative, betzel2017modular}. The resulting group-representative network has the same number of binary connections as the average individual and the same edge length distribution. This type of non-uniform consistency thresholding has been shown to be superior to other, more commonly used forms of thresholding \cite{roberts2017consistency}.\\

\subsection*{Gene coexpression data collection and preprocessing}

\noindent To study conserved genetic features that may constrain ECoG network dynamics, we constructed a correlation matrix of brain regions' gene expression profiles using a similar approach and following \citep{betzel2017inter}. We used normalized microarray data available from the Allen Brain Institute (\url{http://human.brain-map.org/static/download}) \cite{jones2009allen, hawrylycz2012anatomically, sunkin2013allen}. The full dataset includes six donor brains (aged 18 to 68 years) for which spatially-mapped microarray data were obtained ($\approx$ 60,000 RNA probes). We focused on donors \verb|10021| and \verb|9861| which included samples (893 and 946 sites, respectively) from both the left and right hemispheres. Subsequently, we retained only those samples in the cerebral cortex \citep{anderson2018gene}. Next, we extracted expression profiles for each sample, averaged over duplicate genes, and standardized expression levels across samples as $z$-scores. The standardized measure of any sample, then, measured to what extent a particular gene was differentially expressed at that cortical location relative to the other cortical locations in both hemispheres.\\

\noindent In addition to microarray data, the Allen Brain Institute also provides coordinates representing the location in MNI space where each sample was collected. This information facilitated the mapping of sample sites to brain regions in a procedure exactly analogous to our approach for mapping ECoG electrodes. As a result, we obtained representative expression profiles for each brain region (provided there were nearby samples). For each of the two donor brains, we calculated the region-by-region correlation matrix of standarized expression profiles. Due to the overall density of the whole-brain sampling, we were able to generate an estimate of gene expression correlation (a measure of similarity) for 6286 of 6441 possible region pairs ($\approx$97.6\%).\\

\subsection*{Gene-ECoG optimization procedure}

\noindent Following \citep{betzel2017inter}, we identified a subset of genes that were related to resting state ECoG functional connectivity in a separate cohort of patients and from clips far from any ictal activity. Then, we used this subset of genes to construct the gene coexpression matrix utilized in our models. In general, we sought the list of $K$ genes, $\Gamma^{K} = \{g_1,\ldots,g_K \}$ whose brain-wide coexpression matrix was maximally correlated with resting state ECoG functional connectivity in this separate cohort. While the exact solution of this optimization problem is computationally intractable (the full list included 29130 genes), we could define an objective function and use numerical methods to obtain an approximate solution.\\

\noindent The objective function we sought to minimize was defined as follows. Let $G_1(\Gamma)$ and $G_2(\Gamma)$ be the gene coexpression matrices for each of the two donor brains calculated using the gene list, $\Gamma$. We can then vectorize each matrix by extracting its upper triangle of non-zero elements. Then, after doing the same for the resting state ECoG functional connectivity matrix from the separate cohort, $\mathbf{A}^{\text{ECoG}}$, we can calculate the correlation of gene expression with ECoG functional connectivity, resulting in two correlation coefficients $\rho_1$ and $\rho_2$. In general, we wish for the magnitudes of $\rho_1$ and $\rho_2$ to be as large as possible. Accordingly, we defined our objective function to be $F(\rho_1, \rho_2) = \min(\rho_1,\rho_2)$, so that the correspondence of any gene list, $\Gamma$, with ECoG functional connectivity is only as good as the worse of the two donor brains' correlations.\\

\noindent As noted earlier, optimizing this function is computationaly intractable, so we used a simulated annealing algorithm to generate estimates of the solution. In general, simulated annealing works by proposing initial estimates of the solution (that are usually poor), making small changes to these estimates and evaluating whether or not these changes improve the estimate. The algorithm begins in a ``high temperature'' phase, during which even changes that result in inferior estimates can be accepted, making it possible to explore the landscape of possible solutions. Gradually, a temperature parameter is reduced so that in later phases only solutions that result in improvements are accepted.\\

\noindent In our case, the algorithm was initialized with a temperature of $t_0 = 2.5$ and a randomly-generated list of $K$ genes, $\Gamma$, which represented our initial estimate of the solution. From this list we constructed matrices $G_1(\Gamma)$ and $G_2(\Gamma)$, calculated $\rho_1$ and $\rho_2$, and then evaluated the objective function, $F(\rho_1,\rho_2)$. With each iteration, the temperature was reduced slightly, ($t_i = t_{i - 1} \times 0.99975$) and one gene randomly selected from $\Gamma$ was replaced with a novel gene. We then used this new list, $\Gamma^{\prime}$, to construct $G_1(\Gamma)^{\prime}$ and $G_2(\Gamma)^{\prime}$, from which we eventually obtained a new value of the objective function, $F(\rho_1^{\prime},\rho_2^{\prime})$. If $F(\rho_1^{\prime},\rho_2^{\prime}) > F(\rho_1,\rho_2)$, then we replaced $\Gamma$ with $\Gamma^{\prime}$ and the algorithm proceeded to the next iteration. Otherwise, we accepted the $\Gamma^{\prime}$ with probability $\exp(-\frac{[F(\rho_1,\rho_2)-F(\rho_1^{\prime},\rho_2)^{\prime}]}{t_i})$, where $t_i$ is the temperature at the current iteration. The algorithm continued for either 200000 total iterations or 10000 consecutive iterations with no change in $\Gamma$.\\

\noindent The result of simulated annealing will usually vary somewhat from run to run. Accordingly, we repeated the algorithm 50 times. We also varied the number of genes, $K$, from 10 to 360 in increments of 10. We chose the optimal $K$ to be the value at which the objective function was on average greatest over the 50 repetitions. Rather than treat any of the 50 estimated solutions as representative, we calculated how frequently each gene appeared across the ensemble of all 50 solutions, and we compared this frequency to what we would expect in 50 samples of $K$ genes. We retained only those genes that appeared more frequently than expected (false discovery rate controlled at $q = 0.05$). These genes represented the ``optimized list,'' from which we constructed the gene coexpression matrix \textbf{G} = $[\text{G}_{ij}]$, given by the Pearson correlation between the gene expression profiles of region $i$ and region $j$.\\

\subsection*{Generation of Geometric and Structural Network Features}

In this section, we provide formal definitions for the structural network features that we utilized in our multi-linear regression models: \textbf{D} = $[\text{D}_{ij}]$, the Euclidean distance between region $i$ and region $j$; \textbf{G} = $[\text{G}_{ij}]$, the Pearson correlation between gene expression profiles of region $i$ and region $j$; \textbf{S}=$[\text{S}_{ij}]$, the search information between region $i$ and region $j$, a measure of the ``hiddenness'' of the shortest anatomical path between regions; \textbf{P}=$[\text{P}_{ij}]$, the length of the true shortest path length between region $i$ and region $j$; and \textbf{F}=$[\text{F}_{ij}]$, the maximum flow between region $i$ and region $j$, a measure that treats edge weights as capacities and sends units of flow from region $i$ to region $j$ while respecting maximum capacities.

\paragraph{Euclidean Distance.} To estimate the $N \times N$ matrix \textbf{D} of Euclidean distances, we considered the center of mass (COM) of each region of interest in the whole brain parcellation, and we calculated the 3-dimensional Euclidean distance between all possible pairs of COMs.

\paragraph{Path Length.} To estimate the $N \times N$ matrix \textbf{P}=$[\text{P}_{ij}]$ of path lengths, we calculated the shortest path length between region $i$ and region $j$ in the group-representative structural adjacency matrix $\mathbf{A}$ using Dijkstra's algorithm.

\paragraph{Search information.} Anatomical connectivity matrices obtained from diffusion imaging data and reconstructed using deterministic tractography are usually sparse, meaning that only a fraction of all possible connections exist \cite{iturria2008studying, hagmann2008mapping}. Rather than only use the sparse connectivity matrix to predict ECoG functional connectivity, we also generated a full matrix, $S$, whose element $S_{ij}$ indicates the information (in bits) required to follow the shortest path from node $i$ to node $j$ \cite{rosvall2005searchability}. Let $\pi_{s \rightarrow t} = \{ A_{si}, A_{ij}, \ldots , A_{kt} \}$ be the series of structural edges that are traversed along the shortest path from a source node, $s$, to a different target node, $t$, and $\Omega_{s \rightarrow t} = \{s,i,j,\ldots,k,t\}$ be the sequence of nodes along the same path. The probability of following this path under random walk dynamics is given by $P(\pi_{s \rightarrow t}) = \prod_{i \in \Omega^*_{s \rightarrow t}} \frac{\pi_{i \rightarrow t}^{(1)}}{s_i}$, where $s_i = \sum_j A_{ij}$ is the weighted degree of node $i$, $\pi_{i \rightarrow t}^{(1)}$ is the first edge on the shortest path from $i$ to $t$ and $\Omega^*_{s \rightarrow t} = \{s,i,j,\ldots,k\}$ is the shortest path node sequence excluding the target node. The amount of information (in bits) required to access this shortest path, then, is given by $S(\pi_{s \rightarrow t}) = \log_2(P(\pi_{s \rightarrow t}))$. We can treat every pair of nodes $\{i,j\}$ as the source and target, respectively, and (provided that there exists a unique shortest-path from node $i$ to node $j$) we can compute $S(\pi_{i \rightarrow j})$ for all such pairs. The resulting matrix, $S$, termed ``search information'', has been shown to partially predict BOLD FC \cite{goni2014resting} and may be modulated in certain neurological disorders \cite{wirsich2016whole}.\\

\paragraph{Maximum Flow.} Search information, while useful, is affected only by the edges along the shortest path. To incorporate information from longer distance paths in the network, we generated a full matrix $F$, whose element $F_{ij}$ indicates the maximum amount of flow that can be sent through the network from node $i$ to node $j$. In this model, the source node $i$ is treated as an infinite source of flow and the weights of each edge from node $s$ to node $t$ are considered to be the capacity of that edge, or the amount of flow that can be sent directly from node $s$ to node $t$ \citep{elias1956note}. Therefore, the value $F_{ij}$ is the maximum amount of flow that can be sent from node $i$ to node $j$ while respecting all capacities in the network. We can treat every pair of nodes $\{i,j\}$ as the source and sink, respectively, and we can compute $F(\pi_{i \rightarrow j})$ for all such pairs. The resulting matrix, $F$, termed ``maximum flow'', has been shown to be of value in modeling information transfer in non-shortest paths \citep{lyoo2018modelling} and has previously been used to model connectivity changes in patients with Alzheimer's disease \citep{yoo2015network, prasad2013flow}.

\bibliography{bib_v1}
\bibliographystyle{ieeetr}

\end{document}